\documentstyle[11pt,a4]{article}
\def\be{\nopagebreak[3]\begin{equation}}
\def\ee{\end{equation}}
\def\ba{\nopagebreak[3]\begin{eqnarray}}
\def\ea{\end{eqnarray}}
\def\nl{\nonumber \\}
\def\ni{\noindent}
\newcommand{\scs}[1]{{\scriptscriptstyle #1}}
\begin{document}
\begin{titlepage}

\title{
A MODEL OF THREE-DIMENSIONAL LATTICE GRAVITY}
\author{
 D.V. BOULATOV\thanks{Address after 1 October 1992: The Niels Bohr
 Institute, Blegdamsvej 17, DK-2100, Kopenhagen \O , Denmark.}\\
Service de Physique Th\'{e}orique de Saclay\thanks{
Laboratoire de la Direction des Sciences de la Mati\`{e}re du
Commissariat \`{a} l'Energie Atomique},
\and
F-91191 Gif-sur-Yvette Cedex, France}
\date{February 1992}

\maketitle

\vspace{3cm}
\begin{abstract}
\normalsize
A model is proposed which generates all oriented
$3d$ simplicial complexes weighted
with an invariant associated with a topological lattice gauge theory.
When the gauge group is $SU_q(2)$, $q^n=1,$ it is
the Turaev-Viro invariant and
the model may be regarded as a non-perturbative definition of $3d$
simplicial quantum gravity. If one takes a finite abelian group $G$,
the corresponding invariant gives the rank of the first cohomology group
of a complex \nolinebreak $C$: $I_G(C) = rank(H^1(C,G))$,
which means a topological
expansion in the Betti number $b^1$. In general, it is a theory of the
Dijkgraaf-Witten type, $i.e.$ determined completely by the fundamental
group of a manifold.
\end{abstract}

\vfill
\ni
{\small submitted to Physics Letters B} \hfill
SPhT/92-017
\end{titlepage}

\section{Introduction}

Lattice models have always been an useful tool in field theory. They
often helped to look at a theory from another point of view, which led
to better understanding and computational progress. The most recent
example of such a kind was the matrix models of 2d gravity \cite{MatMod}.
In that case lattice and continuum approaches have been developed in
the close connection stimulating each other. The success of the matrix
models made it desirable to extend this approach to higher dimensional
euclidean gravity.
The general idea is rather natural: the integral over all d-dimensional
manifolds should be substituted by a sum over all d-dimensional
simplicial complexes. If a topology is fixed, a lattice
action may be chosen linear in the number of simplexes of
every dimension. The partition function in the $3d$ case is of the form

\be
Z_{top}=\sum_{C_{top}}e^{\alpha N_1 - \beta N_3}
=\sum_{N_1,N_3}Z_{N_1N_3}e^{\alpha N_1 - \beta N_3}
\label{Ztop}
\ee

\ni
where $top$ means a fixed topology, $\sum_{C_{top}}$ is the sum over all
$3d$ simplicial manifolds of the chosen topology. Let us remind that in
odd dimensions manifolds have the zero Euler character, hence

\be
\chi = N_0-N_1+N_2-N_3 = 0
\label{chi}
\ee

\ni
$N_k$ is the number of simplexes of the $k$-th dimension in a complex
$C$, $i.e.$ points, links, triangles and tetrahedra, respectively. The
other constraint is

\be
N_2=2N_3
\ee

\ni
which means simply that every triangle is shared by exactly two
tetrahedra.
Owing to the constraints, if the volume is fixed, only one
parameter remains in the 3d and 4d cases and one may hope that it should
be related to a bare Newton coupling. Indeed, keeping all
tetrahedra equilateral, one gets, from counting deficit  angles associated
with links, the lattice analog of the mean curvature \cite{Regge}

\be
\int d^3x\sqrt{g}R \sim a\bigg(2\pi N_1-6N_3\cos^{-1}\bigg(\frac{1}{3}
\bigg)\bigg)
\label{curvature}
\ee

\ni
where $a$ is a lattice spacing.

For the fixed spherical topology, a
3-dimensional model of such a type
was investigated numerically in refs. \cite{AgMig,BK} and a 4-dimensional
one, in refs. \cite{4dsim}. It appears that the micro-canonical
partition function $Z_{N_1N_3}$ is exponentially bounded at large $N_3$

\be
Z_{N_1N_3} \sim e^{\beta^*N_3}
\ee

\ni
while with respect to $N_1$ at $N_3$ fixed its shape is,
roughly  speaking, gaussian \cite{BK}. Varying the lattice analog of the
inverse Newton constant, $\alpha$, one can only shift the position of
the maximum changing continuously the mean curvature
(\ref{curvature}). It appeared that the vacuum is not unique.
In refs. \cite{BK} the first order phase
transition was found at some $\alpha_c>0$ which separates phases of positive
($\alpha>\alpha_c$) and negative ($\alpha<\alpha_c$) mean curvatures
(\ref{curvature}). The remarkable feature of the first phase is that the
mean curvature per unit volume, $2\pi N_1/N_3-6\cos^{-1}\big(\frac{1}{3}\big)$
, does not depend on $N_3$ at all
\cite{BK}. It means the existence of the continuum thermodynamical
limit for the model (\ref{Ztop}).
A similar transition also exists in the 4-d model
and there is
some hope that here it is of the second order \cite{4dsim}.
If it is confirmed, one
can find a non-trivial continuum limit in its vicinity.
Anyhow, the lattice models of gravity are interesting in their own
rights.

The aim of this paper is to construct a model which
generates all 3-dimensional simplicial complexes within a perturbation
expansion so that it might be regarded as a 3-d analog of the matrix
models. The naive generalization, so-called tensor models \cite{TensorMod},
suffers from serious diseases. The main one is that they do not
contain the sufficient number of parameters: it is impossible to perform
any topological expansion within them.
It makes these models uninteresting because of their
non-universality. As we learnt from the matrix models, only a perturbation
topological expansion might be universal \cite{Matmod2}.
So, one has somehow to control the topology.

It is well-known
that, in the 3d space-time, integration over diffeomorphisms and local
Lorentz rotations is equivalent to the $ISO(2,1)$ Chern-Simons field
theory
\cite{Witten}. Although that connection holds only on-shell, it is
clear that, in general, every topology should be somehow weighted.
So far the Turaev-Viro $SU(2)$ invariant has been considered
as the lattice counterpart of the $ISO(2,1)$ Chern-Simons partition
function \cite{TV,Ooguri}.
Strictly speaking, the corresponding argumentation is heuristic\footnote{For
example, in $ISO(2,1)$ Chern-Simons theory there is no reason to quantize
the coupling constant $k$ in contrast  with the $SU(2)$ case.} and it
may be better to consider as general class of models as
possible. As will be shown in this paper, the underlying structure of the
Turaev-Viro \cite{TV} (or Ponzano-Regge in its original form \cite{PR})
partition function
is 3-dimensional topological lattice gauge theory, and, simply taking different
gauge groups, one is able to construct different invariants.
This "degree
of freedom" appears to be rather useful and, as we hope, will lead to
the better understanding of the problem of $3d$ gravity.

The paper is organized as follows. In Section 2, a model is formulated
which generates all $3d$ simplicial complexes weighted with the
Ponzano-Regge ({\em i.e.} non-regularized) partition function.
{}From the point of view of the Regge calculus \cite{Regge},
this partition function
corresponds to a discretization of $3d$ euclidean gravity \cite{PR}.
A natural generalization leads to a whole class of models of such a type.
In Section 3, the $Z_n$ gauge group is considered. It is shown that, in
some scaling limit, an expansion in the Betti number $b_2$ can be
performed. In Section 4, the case of $q$-deformed $SU(2)$ gauge group is
considered, when the Ponzano-Regge construction leads to the
Turaev-Viro invariant. Section 5 is devoted to a discussion.

\section{General construction}

The basic object is a set of real functions of 3 variables $\phi(x,y,z)$
(where $x,y,z \in G$ for some compact group $G$) invariant under
simultaneous right shifts of all variables by $u \in G$.

\be
\phi(x,y,z) = \phi(xu,yu,zu);\ \ \ \
\overline{\phi}(x,y,z) = \phi(x,y,z)
\label{defofphi}
\ee
\ni
We also demand the cyclic symmetry

\be
\phi(x,y,z) = \phi(z,x,y)= \phi(y,z,x)
\label{cycsim}
\ee
\ni
The general Fourier decomposition of such a function is of the form

\[
\phi(x,y,z) = \sum_{j_1,j_2,j_3} \sum_{\{m,n,k\}}
\Phi^{m_1m_2m_3;k_1k_2k_3}_{\;j_1,\;j_2,\;j_3}
D^{j_1}_{m_1,n_1}(x) D^{j_2}_{m_2,n_2}(y) D^{j_3}_{m_3,n_3}(z)
\]\be
\int d\omega D^{j_1}_{n_1k_1}(\omega) D^{j_2}_{n_2k_2}(\omega)
D^{j_3}_{n_3k_3}(\omega)
\label{phiexp}
\ee
\ni
$n_i,m_i,k_i=1,\ldots,d_{j_{i}}$
($d_j$ is the dimension of an irrep $j$);
$D^j_{nm}(x)$ are matrix elements obeying the
orthogonality condition

\be
\int dx D^j_{nm}(x) \overline{D}^{j'}_{n'm'}(x) = {1\over d_j}\delta^{j,j'}
\delta_{n,n'}\delta_{m,m'}
\label{orthcon}
\ee

\ni
Throughout the paper all
measures are assumed to be normalized to the unity:

\be
\int_G dx \equiv \frac{1}{rank(G)}\sum_{g\in G}1 = 1
\label{normaliz}
\ee

The integral of three matrix elements is proportional to a product of two
Clebsch-Gordan
coefficients $\langle j_1j_2n_1n_2 \mid j_1j_2j_3n_3 \rangle$.
We shall use the following notation:

\newcommand{\threej}[6]{\left( \begin{array}{ccc}
#1&#3&#5 \\ #2&#4&#6
\end{array} \right)}

\newcommand{\sixj}[6]{\left\{ \begin{array}{ccc}
#1&#2&#3 \\ #4&#5&#6
\end{array} \right\}}

\be
\int dx\, D^{j_1}_{m_1n_1}(x)D^{j_2}_{m_2n_2}(x)D^{j_3}_{m_3n_3}(x)
=\threej{j_1}{m_1}{j_2}{m_2}{j_3}{m_3}\threej{j_1}{n_1}{j_2}{n_2}{j_3}{n_3}
\label{int3me}
\ee

\ni
In the $SU(2)$ case,
$\threej{j_1}{n_1}{j_2}{n_2}{j_3}{n_3}$ is called the Wigner $3j$-symbol:

\be
\threej{j_1}{n_1}{j_2}{n_2}{j_3}{n_3}={(-1)^{j_1-j_2+n_3}
\over \sqrt{2j_3+1}}\langle j_1j_2n_1n_2
\mid j_1j_2j_3-n_3 \rangle
\label{def3j}
\ee
\ni
The Fourier coefficients,

\[
A^{m_1,m_2,m_3}_{\;j_1,\;j_2,\;j_3}=
\frac{1}{\sqrt{(2j_1+1)(2j_2+1)(2j_3+1)}} \]\be
\sum_{k_1k_2k_3}
\Phi^{m_1m_2m_3;k_1k_2k_3}_{\;j_1,\;j_2,\;j_3}
\threej{j_1}{k_1}{j_2}{k_2}{j_3}{k_3}
\ee
\ni
are complex numbers having symmetries of $3j$-symbols except for
the condition

\be
m_1+m_2+m_3=0
\ee

An action of interest can be constructed with those functions as
follows

\ba
&&S =\frac{1}{2}\int dxdydz\, \phi^2 (x,y,z)-\nl &&
- \frac{\lambda}{4!} \int dxdydzdudvdw\;
\phi(x,y,z)\phi(x,u,v)\phi(y,v,w)\phi(z,w,u)
\label{action1}
\ea

\ni
If the variables are attached to edges, the first term can be regarded
as two glued triangles and the second, as four triangles forming a
tetrahedron.
Integrating out all group variables, one gets in the $SU(2)$ case

\ba
&&S = \frac{1}{2}\sum_{\{j_1,j_2,j_3\}}\sum_{\{-j_k \le m_k \le j_k\}}
\mid A_{\;j_1,\;j_2,\;j_3}^{m_1,m_2,m_3} \mid^2 -
\nl&&
- \frac{\lambda}{4!}\sum_{\{j_1,...,j_6\}}\sum_{\{-j_k \le m_k \le j_k\}}
(-1)^{\sum_k^6 (m_k+j_k)}
A_{\ j_1,\ j_2,\ j_3}^{-m_1,-m_2,-m_3}
A_{\;j_3,\ j_4,\;j_5}^{m_3,-m_4,m_5}  \nl&&
A_{\;j_1,\ j_5,\;j_6}^{m_1,-m_5,m_6}
A_{\;j_2,\ j_6,\;j_4}^{m_2,-m_6,m_4}
\sixj{j_1}{j_2}{j_3}{j_4}{j_5}{j_6}
\label{Sintout}
\ea

\ni
If the coefficients $A_{\;j_1,\;j_2,\;j_3}^{m_1,m_2,m_3}$ obey the
condition following from the reality of $\phi(x,y,z)$
(eq. (\ref{defofphi})),

\be
\overline{A}_{\;j_1,\;j_2,\;j_3}^{m_1,m_2,m_3} =
(-1)^{\sum_i^3(m_i+j_i)}A_{\ j_1,\ j_2,\ j_3}^{-m_1,-m_2,-m_3}
\label{conjA}
\ee

\ni
and the measure of integration is taken to be

\be
{\cal D}\phi = \prod_{\{j,j',j''\}}\prod_{\{-j \le m \le j\}}
dA_{\;j,\;j',\;j''}^{m,m',m''}
\ee

\ni
where $\prod_{\{j,j',j''\}}$ means the product over all triplets $(j,j',j'')$
obeying
the triangle inequality: $\mid j' - j''\mid \le j \le j' + j''$, then
the partition function will generate all possible $3d$ simplicial
complexes weighted with corresponding (non-regularized)
Ponzano-Regge partition functions, {\it i.e.}

\be
Z = \int {\cal D}\phi\; e^{-S} =
 \sum_{\{C\}} \lambda^{N_3(C)}
\sum_{\{j\}} \prod_{L \in C} (2j_{\hbox{}_L}+1)
\prod_{T \in C}
\sixj{j_{\hbox{}_{T_1}}}{j_{\hbox{}_{T_2}}}{j_{\hbox{}_{T_3}}}
{j_{\hbox{}_{T_4}}}
{j_{\hbox{}_{T_5}}}{j_{\hbox{}_{T_6}}}
\label{Z}
\ee

\ni
where $\sum_{\{C\}}$ is the sum over all oriented $3d$ simplicial
complexes;
$N_3(C)$ is the number of tetrahedra in a complex $C$, $\sum_{\{j\}}$
is the sum over all possible configurations of $j$'s
(colorings of links); $\prod_{\{L \in
C\}}$ is the product over all links $L$ in $C$; $\prod_{\{T \in C\}}$ is
the product over all tetrahedra $T$ ($j_{\hbox{}_{T_i}}, i=1,\ldots,6$;
are six momenta attached to edges of a tetrahedron $T$).
$\sixj{j_{\hbox{}_{T_1}}}{j_{\hbox{}_{T_2}}}{j_{\hbox{}_{T_3}}}
{j_{\hbox{}_{T_4}}}
{j_{\hbox{}_{T_5}}}{j_{\hbox{}_{T_6}}}$
is the Racah-Wigner $6j$-symbol attached to a tetrahedron $T$.
We use the normalization for which the $6j$-symbol is symmetric with
respect to permutations of columns:

\[
\sixj{j_1}{j_2}{j_3}{j_4}{j_5}{j_6}=
\sum_{\{-j_i\leq m_i \leq j_i\}} (-1)^{j_4+j_5+j_6+m_4+m_5+m_6}
\threej{j_1}{m_1}{j_2}{m_2}{j_3}{m_3}
\]\be
\threej{j_5}{m_5}{j_6}{-m_6}{j_1}{m_1}
\threej{j_6}{m_6}{j_4}{-m_4}{j_2}{m_2}
\threej{j_4}{m_4}{j_5}{-m_5}{j_3}{m_3}
\ee

Eq.(\ref{Z}) is formal and has to be somehow regularized. Let us
postpone a discussion on that and firstly make several remarks.
Eq.(\ref{defofphi}) means that we are considering functions of two
independent variables. If we drop the cyclic symmetry condition
(\ref{cycsim}), then we shall have the representation

\[
\phi(x,y,z) = f(xz^+,yz^+) = \sum_{j_1;m_1,n_1}\sum_{j_2;m_2,n_2}
F^{m_1,n_1;}_{\ j_1} \hbox{}^{m_2,n_2}_{\ j_2}
 D^{j_1}_{m_1n_1}(xz^+) D^{j_2}_{m_2n_2}(yz^+)  \]\[
=\sum_{j_1;m_1,n_1}\sum_{j_2;m_2,n_2}\sum_{j_3;m_3,n_3}
F^{m_1,n_1;}_{\ j_1} \hbox{}^{m_2,n_2}_{\ j_2}
\langle j_1 j_2 m'_1 m'_2 \mid j_1j_2j_3-\!\!n_3 \rangle\]\be
\langle j_1 j_2 n_1 n_2 \mid j_1j_2j_3-\!\!m_3 \rangle
(-1)^{m_3+n_3}
 D^{j_1}_{m_1{m'}_1}(x) D^{j_2}_{m_2{m'}_2}(y)
 D^{j_3}_{m_3n_3}(z)
\ee

\ni
Hence,

\be
\widetilde{A}_{\;j_1,\;j_2,\;j_3}^{m_1,m_2,m_3} =
\sum_{n_1,n_2}\sqrt{\frac{(2j_3+1)}{(2j_1+1)(2j_2+1)}}
\threej{j_1}{n_1}{j_2}{n_2}{j_3}{m_3}
F^{m_1,n_1}_{\ j_1}\, \hbox{}^{m_2,n_2}_{\ j_2}
\ee

\ni
{}From eq.(\ref{conjA}) it follows that

\be
\overline{F}^{m_1,n_1;}_{\ j_1} \hbox{}^{m_2,n_2}_{\ j_2}=
(-1)^{m_1+m_2+n_1+n_2}
F^{-\!m_1,-\!n_1;}_{\ j_1} \hbox{}^{-\!m_2,-\!n_2}_{\ j_2}
\label{conjF}
\ee
\ni
and, if the correlator of the Fourier coefficients is of the form

\ba
&\langle F^{m_1,n_1;}_{\ j_1} \hbox{}^{m_2,n_2}_{\ j_2}
\overline{F}^{m'_1,n'_1;}_{\ j'_1} \hbox{}^{m'_2,n'_2}_{\ j'_2}\rangle =
(2j_1+1)(2j_2+1) \delta_{j_1,j'_1}\delta_{j_2,j'_2}&\nl
&\delta^{m_1+m'_1,0}\delta^{m_2+m'_2,0}
\delta^{n_1+n'_1,0}\delta^{n_2+n'_2,0}&
\label{corrF}
\ea
then

\[
\langle
\widetilde{A}_{\;j_1,\;j_2,\;j_3}^{m_1,m_2,m_3}
\overline{\widetilde{A}}_{\;j'_1,\;j'_2,\;j'_3}^{m'_1,m'_2,m'_3}\rangle =
\sum_{n_1,n_2}
\threej{j_1}{n_1}{j_2}{n_2}{j_3}{m_3}
\threej{j_1}{n_1}{j_2}{n_2}{j'_3}{-m'_3}\]\nopagebreak\be
\delta_{j_1,j'_1}\delta_{j_2,j'_2}
\delta^{m_1+m'_1,0}\delta^{m_2+m'_2,0}
=\delta_{j_1,j'_1}\delta_{j_2,j'_2}\delta_{j_3,j'_3}
\delta^{m_1+m'_1,0}\delta^{m_2+m'_2,0}\delta^{m_3+m'_3,0}
\ee

\ni
In terms of the function $f(x,y)$ the action (\ref{action1}) takes the
form

\ba
&&S =\frac{1}{2}\int dxdy\, f^2 (x,y) \nl &&
-\frac{\lambda}{4!} \int dxdydudvdw\; h(x,y)h(xw,uw)h(v,u)h(vw,yw) \nl &&
\label{action2}
\ea
\ni
where $h(x,y)=\frac{1}{3}(f(x,y)+f(yx^+,x^+)+f(y^+,xy^+))$.

For a general compact group, the action (\ref{action2}), as well as
(\ref{action1}), together with
the reality condition

\be
\overline{f}(x,y) = f(x,y)
\ee

\ni
may be regarded as a definition of the model. The underlying
mathematical structure here is topological lattice gauge theory. It
can be seen as follows. The action (\ref{action1}) generates $3d$
complexes so that two $3j$-symbols are attached to every triangle.
Such a combination can be
obtained integrating three matrix elements as in eq. (\ref{int3me}).
All lower indices of the matrix elements are sumed up
inside tetrahedra forming $6j$-symbols.
It is easy to notice that
the partition function (\ref{Z}) can be written then in the form

\[
Z=\sum_{\{C\}} \lambda^{N_3}\!\!\!\!\sum_{\{j;n,m\}}\prod_{\{L\in C\}}
d_{j_{\hbox{}_L}}\prod_{t\in C}
\int dx_t\, D^{j_{t_1}}_{m_{t_1}n_{t_1}}(x_t)D^{j_{t_2}}_{m_{t_2}n_{t_2}}(x_t)
D^{j_{t_3}}_{m_{t_3}n_{t_3}}(x_t)\]\[
=\sum_{\{C\}} \lambda^{N_3}\int\prod_{t\in C}dx_t
\prod_{\{L\in C\}}\delta(\prod_{around\,L}x_{t_{\hbox{}_L}},1)\]
\be=\sum_{\{C\}} \lambda^{N_3}Z_{gauge}(C)
\label{Z2}
\ee

\ni
where $\prod_{\{L\in C\}}$ and $\prod_{\{t\in C\}}$ are products over all
links and triangles, respectively. The matrix elements being multiplied
around links produce characters and then, summing over representations,
one gets a $\delta$-function for every link. Its argument,
$\prod_{around\,L}x_{t_{\hbox{}_L}}$, are the product of group elements
$x_{t_{\hbox{}_L}}$ around a link $L$. Triangles are oriented, the
change of an orientation leading to the conjugasion: $x\to x^+=x^{-1}$.
All products have to be performed taking the orientation into account.
Although it is not the fact for a general compact group, in the $SU(2)$
case our model generates only oriented complexes. It follows immediately
from (\ref{conjA}) (or (\ref{conjF})) and the form of the action
(\ref{Sintout}).

If a complex is fixed, the model is equivalent to $3d$ gauge theory with
fields defined on links of the dual $\phi^4$ graphs and the pure gauge
condition on dual faces. If $\delta$-functions were substituted by, for
example, the heat-kernel weights

\be
\delta(x)=\sum_jd_j\chi_j(x) \leftarrow
{\cal G}(x)=\sum_jd_j\chi_j(x) e^{-\epsilon C_j}
\label{heatkernel}
\ee

\ni
where $C_j$ is the quadratic Casimir, one would have just the ordinary
lattice gauge theory. The former is the weak coupling limit of the
latter.

Now, to prove the topological invariance, we need only formal properties
of the group measure and $\delta$-functions. We have to investigate
transformations of $Z_{gauge}(C)$ under topology preserving deformations of
a complex $C$. Two complexes are of the same topological type
(homeomorphic), if they can be connected by a sequence of elementary
"continuous" deformations (moves). These moves
can be defined as follows \cite{AgMig,BK}: if some subcomplex of a
$d$-dimensional complex
can be identified with a part of the boundary of the
$(d+1)$-dimensional simplex, it is substituted by the rest of the boundary.
In the 3-dimensional case there are two pairs of mutually inverse moves
shown in Fig. 1 (a) and (b). The first pair is called the triangle-link
exchange (the dual diagrams are shown in Fig. 2(a)). A pair of tetrahedra
glued together by faces  is substituted by three tetrahedra sharing the
new link. For the first configuration we have the integral of the type

\ba
&\int dx_1dx_2dx_3dy_1dy_2dy_3dw D^{j_1}_{m_1n_1}(x_1x_2^+)
D^{j_2}_{m_2n_2}(x_2x_3^+)D^{j_3}_{m_3n_3}(x_3x_1^+)&\nl&
D^{j'_1}_{m'_1n'_1}(y_1y_2^+)D^{j'_2}_{m'_2n'_2}(y_2y_3^+)
D^{j'_3}_{m'_3n'_3}(y_3y_1^+)&\nl&
D^{l_1}_{a_1b_1}(x_1wy_1^+)D^{l_2}_{a_2b_2}(x_2wy_2^+)
D^{l_3}_{a_3b_3}(x_3wy_3^+)&
\label{2tetr}
\ea

\ni
$x$'s and $y$'s stay for faces of the upper and lower tetrahedra,
respectively, and $w$, for the common face. It is clear that the
dependence on $w$ can be removed by the shift $x_1\to x_1w^+$; $x_2\to
x_2w^+$; $x_3\to x_3w^+$. In the second case the situation is quite
analogous: there are three triangles ($w$'s) and one link
($\delta$-function) inside the subcomplex. The
counterpart of eq. (\ref{2tetr}) is

\ba
&\int dx_1dx_2dx_3dy_1dy_2dy_3dw_1dw_2dw_3D^{j_1}_{m_1n_1}(x_1w_1x_2^+)
D^{j_2}_{m_2n_2}(x_2w_2x_3^+)&\nl&D^{j_3}_{m_3n_3}(x_3w_3x_1^+)
D^{j'_1}_{m'_1n'_1}(y_1w_1y_2^+)D^{j'_2}_{m'_2n'_2}(y_2w_2y_3^+)
D^{j'_3}_{m'_3n'_3}(y_3w_3y_1^+)&\nl&
D^{l_1}_{a_1b_1}(x_1y_1^+)D^{l_2}_{a_2b_2}(x_2y_2^+)
D^{l_3}_{a_3b_3}(x_3y_3^+)\delta(w_1w_2w_3,1)&
\label{3tetr}
\ea

\ni
where all $w$-integrations are trivial due to the $\delta$-function.

Instead of proving the invariance under the moves in Fig.1(b), we can
consider the case of two tetrahedra glued along three faces
(Fig.2(b)). This configuration can be obtained removing one of the links in
Fig.1(b) by the triangle-link exchange. The appearing integral is of
the form

\ba
&\int dw_1dw_2dw_3\, \delta(w_1w_2^+)\delta(w_2w_3^+)\delta(w_3w_1^+)
D^{j_1}_{m_1n_1}(w_1)D^{j_2}_{m_2n_2}(w_2)D^{j_3}_{m_3n_3}(w_3)&\nl&
=\delta(1,1)\int dwD^{j_1}_{m_1n_1}(w)D^{j_2}_{m_2n_2}(w)
D^{j_3}_{m_3n_3}(w)
\label{pointout}
\ea

\ni
which means that, up to $\delta(1,1)=rank(G)$, those two glued tetrahedra
are equivalent to a single triangle. We see that the partition
function (\ref{Z}) can be finally written down as

\be
Z=\sum_{\{C\}} \lambda^{N_3}(rank(G))^{N_0-1}I_{\hbox{}_G}(C)
\label{Z3}
\ee

\ni
where $I_{\hbox{}_G}(C)$ is a topological invariant associated with a
group $G$.

For finite groups, our model is well defined, as in this case the rank
is equal to the number of group elements. For continuous compact groups,
the $q$-deformation provides us with a regularization of the model
(notice that the substitution (\ref{heatkernel}) destroys
the topological properties of the gauge theory). For example, in the
$SU_q(2)$ case, $q^n=1$,

\be
rank(SU_q(2))=\frac{n}{2\sin^2(\frac{\pi}{n})}
\label{rankSU(2)}
\ee

\ni
Indeed, the representations of the $q$-analogs of
compact groups resemble the classical representations. And, while one is
working with $3j$ and $6j$-symbols not permuting momenta,
as we did above, the $R$-matrix does not appear and all formal manipulations
coincide in both cases\footnote{A more complete discussion will be given
in Section 4.}.
We see that quantum groups are here
on equal footing with finite groups.
That is why in the next section we shall
concentrate ourselves on the simpler latter case.

\section{Topological gauge theory for finite groups and the $Z_p$
model.}

The topological lattice model appeared in the previous section is a
particular example of the Dijkgraaf-Witten
theory \cite{DijkW}. Actually, it is the simplest model of such a type.
Dijkgraaf and Witten introduced a topological action, which exists,
however, not for all groups. In our case there is no action and, therefore,
there are no corresponding restrictions.
Two other peculiarities are (i) the gauge
fields are defined on dual edges rather than on links of a
triangulation; (ii) since $\sum_{\{C\}}$ runs over all possible
complexes, we should take into consideration non-manifolds as well.
Nevertheless, the model bears general properties of the Dijkrgaaf-Witten
one. The main is that its partition function is determined completely by
the fundamental group.

Among lattices generated perturbatively there are such that, strictly
speaking,
do not obey the definition of the simplicial complex (for example, shown
in Fig.2 (b)). It forces us to work with more general cell complexes.

\newcommand{\wtil}[1]{\widetilde{#1}}

{}From now on, we shall consider simultaneously triangulations and dual
$\phi^4$ lattices denoting quantities defined for the latter by the
tilde $\wtil{\ }$. So, at the beginning we have a cell complex
dual to a triangulation: 0-cells are counterparts of tetrahedra, 1-cells
of triangles, 2-cells of links and 3-cells of vertices of the
triangulation. Since analogs of eqs. (\ref{2tetr},\ref{3tetr},\ref{pointout})
are valid for
general polyhedra as well, we can shrink a 1-cell identifying two
0-cells forming its boundary (Fig. 3(a)); delete a 2-cell joining two
3-cells a common boundary of which it was (Fig. 3(b))
and drop a subcomplex
homotopic to a $3d$ spherical ball (Fig. 3(c)). Of course,
all these manipulations are possible only when a final complex
is homotopic to an initial one.
So, we have in hands the powerful apparatus of the cell homology theory.

Given a cell complex $\wtil{C}$, one can easily calculate the corresponding
invariant as follows:\\
1) all complexes under consideration should be put in the form
where there are only one 3-cell $\sigma^3$ and only one 0-cell
$\sigma^0$. It is always
possible for oriented connected manifolds.\\
2) a gauge variable $g_i\in G$ is put into correspondence to every 1-cell
$\sigma^1_i;\ i=1,\ldots,n_1$.\\
3) each 2-cell $\sigma^2_j;\ j=1,\ldots,n_2$; gives a $\delta$-function with
the argument equal to the
ordered product of the gauge variables along its boundary, $\partial
\sigma^2_j$, taking an orientation
into account (the inversion $g\to g^{-1}$ corresponding to the moving in
the opposite direction). If the boundary is empty, one has to
substitute $\delta(1,1)=rank(G)$.\\
Finally, one gets

\be
\wtil{I}_G(\wtil{C})=\int_{G}\prod_{i=1}^{n_1}dg_i
\prod_{j=1}^{n_2}\delta\big(\prod_{\ell\in \partial \sigma^2_j} g_{\ell},1\big)
\label{invariant}
\ee
\ni
$n_1$ and $n_2$ are the numbers of 1-cells and 2-cells, respectively.

Let us point out the simple fact that {\it the $n_2$ conditions

\be
\prod_{\ell\in \partial \sigma^2_j} g_{\ell}=1
\ee
\ni
can be regarded as the defining
relations of the fundamental group $\pi_1(\wtil{C})$, if one
considers $G$ as the free group on $n_1$ generators}.

{}From eq.(\ref{invariant}) it follows that

\be
\wtil{I}_G(\wtil{C})=rank(\pi_1(\wtil{C})\stackrel{h}{\mapsto}G)
\label{invariant2}
\ee
\ni
which is reminiscent of theories of the Dijkgraaf-Witten type.
$\pi_1(\wtil{C})\stackrel{h}{\mapsto}G$ means the homomorphism of
$\pi_1(\wtil{C})$ into a finite group
$G$ defined by the above construction.

{}From eq. (\ref{invariant}) it follows that $\wtil{I}_G(\wtil{C})$ is
multiplicative with respect to the connected sum of two $3d$ complexes,
$\wtil{C}=\wtil{C}_1\#\wtil{C}_2$,

\be
\wtil{I}_G(\wtil{C})=\wtil{I}_G(\wtil{C}_1)\,\wtil{I}_G(\wtil{C}_2)
\ee
\ni
The operation $\#$ is commutative, hence, eq. (\ref{invariant}) can be
regarded as a representation of this semi-group.

An interesting case is abelian groups. Since

\be
H_1(\wtil{C},G)=\pi_1(\wtil{C})/[\pi_1(\wtil{C}),\pi_1(\wtil{C})]
\ee
\ni
($i.e.$ the first homology group is a commutated fundamental group), we
have in this case

\be
\wtil{I}_G(\wtil{C})=rank(H_2(\wtil{C},G))
\label{Iabel}
\ee
\ni
where $H_2(\wtil{C},G)$ is the second homology group of a complex
$\wtil{C}$ with coefficients in $G$.

To prove eq. (\ref{Iabel}) let us note that there are only one 0-cell
$\sigma^0$
and only one 3-cell $\sigma^3$ and for all 1-cells $\sigma^1_i$;
$i=1,\ldots,n_1$

\be
\partial \sigma^1_i=0
\ee
\ni
where $\partial$ is the standard homologic boundary operator ($\partial:
\sigma^k_* \to \sigma^{k-1}_*$).
Because of the orientability,

\be
\partial \sigma^3 =0
\ee
\ni
as well ($i.e.$ there are no exact 2-cells) and, hence, every 2-cell
having zero boundary gives a generator of $H_2(\wtil{C},G)$. But it is
exactly the condition that is coded in the arguments of the
$\delta$-functions in eq.(\ref{invariant}): $rank(H_2(\wtil{C},G))$ is
equal to the number of times the $\delta$-functions "have worked".

The group $H_2(\wtil{C},G)$ is isomorphic to $H^1(C,G)$ by the
Poincar\'{e} duality generated by the transformation from $\phi^4$
graphs to triangulations and $vice$ $versa$.

Eq. (\ref{Iabel}) allows us to determine the Betti number {\bf mod} $G$:

\be
b_1=\bigg[\frac{\log \wtil{I}_G(\wtil{C})}{\log rank(G)}\bigg]
\ee
\ni
where $[x]$ means the integer part of $x$.

Now, let us give several simple examples for the cyclic group $Z_p$.\\
1) $Sphere\ S^3$. There are no 1- and 2-cells at all.

\be
\wtil{I}_G(S^3)=1
\ee
\ni
2) $Lenses\ L^q=S^3/Z_q$. There is one 1-cell and one 2-cell: $\partial
\sigma^2 = q\sigma^1$.

\be
\wtil{I}_G(L^q)=\int_G dg\, \delta(g^q,1)
\ee

\be
\wtil{I}_{Z_p}(L^q)=\left\{\begin{array}{ll}
p&\ \ \ \ ,p=q\\
1&\ \ \ \ ,p\ne q
\end{array}\right.
\ee
\ni
3) $S^1\times S^2$ There is one 1-cell and one 2-cell: $\partial
\sigma^2 = 0$.

\be
\wtil{I}_G(S^1\times S^2)=\int_G dg \delta(1,1) = rank(G)
\ee
\be
\wtil{I}_{Z_p}(S^1\times S^2)=p
\ee
4) $S^1\times M^2_r$ where $M^2_r$ is a $2d$ oriented surface with
$r$ handles $r\ge 1$:

\[
\wtil{I}_G(S^1\times M^2_r)=\int_G dg\prod_{i=1}^r df_idh_i \;
\delta\big(\prod_{j=1}^rh_jf_jh_j^{-1}f_j^{-1},1\big)\]
\be
\prod_{j=1}^r\delta(gh_jg^{-1}h_{j}^{-1},1)
\delta(gf_jg^{-1}f_{j}^{-1},1)
\ee
\be
\wtil{I}_{Z_p}(S^1\times M^2_r)=p^{2r+1}
\ee

The consideration so far involved more or less standard things and now
let us discuss peculiarities. First, we should extend our construction
to non-manifolds. In three dimensions there is no general restriction on
the Euler character but in our case $\chi$ defined by eq. (\ref{chi})
appears to be non-negative. It can be seen as follows. For each vertex
in a complex, tetrahedra touching it form a 3-ball with a non-trivial, in
general, $2d$ boundary. Let us denote $\chi^{(2)}_i=2(1-p_i)$ the $2d$ Euler
character of the boundary of the ball for the $i$-th vertex. Summing
over vertices one gets

\be
\sum_{i=1}^{N_0}\chi^{(2)}_i=2N_0-2\sum_{i=1}^{N_0}p_i
\ee
\ni
On the other hand, this quantity can be obtained counting the numbers of
simplexes of different dimensions. A simple algebra gives

\be
\chi =\sum_{i=1}^{N_0}p_i \ge 0
\ee
\ni
By definition a complex is a manifold, iff $\forall \ i:\ p_i=0$.

The Euler character can be as well expressed through the Betti numbers:

\be
\chi=b_0-b_1+b_2-b_3
\ee
\ni
and, since, for oriented connected complexes, always $b_0=b_3$, we have
the inequality

\be
b_2 \geq b_1
\ee
\ni
The dual quantities, $\wtil{b}_i=b_{3-i}$ and $\wtil{\chi}=-\chi$ by the
Poincar\'{e} duality, which reads $H^k(C,Z_p) = H_{3-k}(\wtil{C},Z_p)$.
Hence, our invariant is sensitive to $b_1$.

For manifolds, in eq. (\ref{invariant}), the number of integrations is
always equal to the number of $\delta$-functions ($n_1=n_2$). In
general, there can be an excess of variables. It means that, at least
for abelian groups, the invariant does not distinguish between manifolds
and non-manifolds.
For every manifold there are infinitely many non-manifolds (having
different $\chi$'s) giving the same answer.
Therefore, the choice $G=Z_p$ looks rather reasonable. The invariant gives
essentially $p^{b_1}$ (up to subtleties clearly seen in the case of lenses).
And, if we weigh links with $\mu/p$, triangles with $p$
and tetrahedra with $\lambda/p$, the partition function
will take the form

\be
\log Z=\sum_{\{C_c\}} Q(C)\lambda^{N_3}\mu^{N_1}p^{b_2-1}
\label{Zfin}
\ee
\ni
where the factor $Q(C)=I_{Z_p}(C)/p^{b_1} < p$;
$\sum_{\{C_c\}}$ is the sum over connected oriented complexes.

So, we arrive at the following
generalization of the $2d$ matrix models

\[
Z=\int \prod_{a,b,c=1}^{\mu/p}\prod_{(i,j,k)}d\phi_{i,j,k}^{abc} \exp \bigg\{
-\frac{1}{2}\sum_{a,b,c=1}^{\mu/p}\sum_{(i,j,k)}\mid\phi_{i,j,k}^{abc}\mid^2
\]\be+\frac{\lambda p}{4!}
\sum_{\stackrel{\scriptstyle a,b,c}{d,e,g=1}}^{\mu/p}
\sum_{\stackrel{\scriptstyle i,j,k}{l,m,n}}
\phi_{i,j,k}^{abc}\phi_{-i,l,-m}^{ade}\phi_{-j,m,-n}^{beg}
\phi_{-k,n,-l}^{cgd}
\bigg\}
\label{Zpmodel}
\ee
\ni
Lower indices, $i,j,k,l,m,n$, are taken ${\rm\bf mod}\ p$;
$\phi^{abc}_{i,j,k}=0$, unless $i+j+k=0\ ({\rm\bf mod}\ p)$, and all sums and
products run over this set of indices.
The field $\phi_{i,j,k}^{abc}$ has to obey the following additional
conditions

\be
\phi_{i,j,k}^{abc}=\phi_{j,k,i}^{bca}=\phi_{k,i,j}^{cab}
\ee
\ni
and

\be
\overline{\phi}_{i,j,k}^{abc}=\phi_{-i,-j,-k}^{abc}
\ee

If $p$ is odd, all complexes generated by the model are oriented and the
above analysis is valid.
In the formal limit $p\to 0$ only homologic spheres survive. But one
should be very careful here. There are infinitely many topologies at
$b_2=0$ (all lenses among them). In our model their number is cut
by the volume, $N_3$. Hence, one should keep $p$ sufficiently large (at
least larger than the biggest $q$ among appearing lenses $L^q$). It
means that, for a given $\lambda$ away from a critical point
$\lambda_c$ ($N_3$ is finite), one should take the limit $p\to \infty$
first. After that one may tend $\lambda \to\lambda_c$ performing
simultaneously an analytical continuation to $p=0$. It means a
non-trivial scaling. In any case, one has somehow to remove a
singularity at $\lambda=0$ like it was done for the matrix models in
refs.
\cite{Matmod2}. The problem, however, is whether the number of complexes
with $b_2$ fixed is exponentially bounded. If it is so, the critical value
$\lambda_c$ exists and the above program is self-consistent. If not,
then a further topological classification is needed. For a fixed
topology, the answer to that question is "yes". At least, numerical
experiments clearly showed that the number of spheres homeomorphic to
$S^3$ grows exponentially with the volume. This growth should be
determined locally, as in the $2d$ case, $i.e.$ independently of a
topology. So, the question is "How many topologies can one fill a given
volume with?". But, even if the above program does make sense, hard
technical problems still remain to be overcome.

\section{The $SU_q(2),\ q^n=1$, model.}

In the case of $q$-deformed $SU(2)$ group, some conceptual problems
still remain. The main tool of our analysis in Section 2  was the
Peter-Weyl theorem stating that the algebra of regular functions on a
compact group is isomorphic to the algebra of matrix elements of finite
dimensional representations. The $q$-analog of this theorem was proved
in refs. \cite{Woron} for $\vert q \vert < 1$. In this case there is
one-to-one correspondence between representations of $SU_q(2)$ and
$SU(2)$, and the notion of the matrix elements is naturally generalized.
The main difference in the quantum case is that the tensor product is
not commutative (for example, $\delta(x,y)\neq\delta(y,x)$).
Although in this case there exists a definition of a rank which appeares
to be finite \cite{Majid}, the lattice topological gauge theory built
with this group does not exist because of divergencies.
$q^n=1$ changes the situation drastically. The analysis of
refs. \cite{Woron} is not valid in this case and the whole subject has to
be revised. On the other hand, the theory of representations of the
quantized universal enveloping algebra ${\cal U}_q(SL(2))$, when
$q^n=1$, was given in refs. \cite{repsu2} and in the most complete form
in ref. \cite{Keller}.

As was established in \cite{Keller}, all highest weight irreps $\rho_j$ of
$\ {\cal U}_q(SL(2))$, when $q^n=1$, fall into two classes:\\
a) dimension of $\rho_j,\ dim(\rho_j)\ <M$, where
$M=\left\{\begin{array}{ll}
n/2&\mbox{,$n$ even}\\
n&\mbox{,$n$ odd}
\end{array}\right.$\\
These irreps are numbered by two integers $d$ and $z$: $\langle
d,z\rangle$, where $d=dim(\rho_j)$, and the highest weights are

\be
j=\frac{1}{2}(d-1)+\frac{n}{4}z
\label{highw}
\ee
\ni
b) $dim(\rho_j)=M$. In this case irreps $I^1_z$ are labeled by a complex
number $z\in {\rm C}\backslash\big\{{\rm Z}+\frac{2}{n}r\mid 1\leq r\leq
M-1\big\}$ and have the highest weights

\be
j=\frac{1}{2}(M-1)+\frac{n}{4}z
\ee

There are also indecomposable representations which are not
irreducible but nevertheless cannot be expanded in a direct sum of
invariant subspaces. They are labeled by an integer $2\leq p\leq M$ and
the complex number $z$: $I^p_z$. Their dimension $dim(I^p_z)=2M$.

In ref. \cite{Keller} the following facts important for us were
established:\\
1) If $n\geq 4$, irreps $\langle d,0\rangle$ are unitary only for even
$n$.\\
2) Representations of the type $I^p_z,\ 1\leq p\leq M$ form a two sided
ideal in the ring of representations ({\em i.e.}, if at least one of
them appears in a tensor product, then all representations in the
decomposition will be of this type). Their
quantum dimension vanishes:
$dim_q(I^p_z)=\left\{\begin{array}{ll}
[M]&,p=1\\
\![2M]&,p\geq 2
\end{array}\right\}=0$,
 where $[x]=\frac{q^x-q^{-x}}{q-q^{-1}}$.\\
3) For the tensor product of two irreps the following formula takes
place:

\[
\langle i,z\rangle\otimes\langle j,w\rangle=
\bigg(\bigoplus_{k=\vert i-j\mid+1;+3;+5,\ldots}
^{\min(i+j-1,2M -i-j-1)}\langle k,z+w\rangle\bigg)\oplus
\]\be
\bigg(\bigoplus_{\ell=r,r+2,r+4,\ldots}^{i+j-M}
I^{\ell}_{z+w}\bigg)
\label{tenprod}
\ee
\ni
where $r=\left\{\begin{array}{ll}
1&\mbox{,$i+j-M$ odd}\\
2&\mbox{,otherwise}
\end{array}\right.$\\

Eq. (\ref{tenprod}) means that the class of representations $\langle
d,z\rangle$
and $I^p_z$ with $z=0$ form a ring with respect to the tensor product.
The highest weights (\ref{highw}) are in the one-to-one correspondence with
the ones at $\vert q\vert<1$. Let us {\em suppose} that, for even $n\geq
4$, matrix elements of the first $n/2-1$ irreps of $SU_q(2)$, $\vert
q\vert<1$, allows a limit $q\to e^{\frac{2\pi i}{n}}$, and form (together
with their descendants) the above mentioned ring. On the other hand, we
can ignore $I^p_z$ representations, while we calculate integrals of
products of matrix elements. Hence, we have to
truncate the space of functions to integrate over  to a subspace spanned
by the matrix elements of irreps of the type $\langle d,0\rangle,\ 1\leq
d\leq n/2-1$, for even $n\geq 4$. Then we have a guarantee that appearing
invariants coincide with the Turaev-Viro ones. This construction reminds
very much the finite-groups one considered in the previous section.

In the quantum case we have to correct a number of formulas of Section
\nolinebreak
2. For a unitary representation we still have

\be
D^j_{nm}(x^{-1})=\overline{D}^j_{mn}(x)
\ee
\ni
but the orthogonality condition (\ref{orthcon}) need to be modified as
follows

\[
\int dx D^j_{nm}(x) \overline{D}^{j'}_{n'm'}(x) = {q^{2m}\over [2j+1]}
\delta^{j,j'}\delta_{n,n'}\delta_{m,m'}
\]\be
\int dx \overline{D}^{j'}_{n'm'}(x) D^j_{nm}(x) = {q^{-2n}\over [2j+1]}
\delta^{j,j'}\delta_{n',n}\delta_{m',m}
\label{orthconq}
\ee

To integrate over $SU_q(2)$ variables in eq. (\ref{action1}), we can use
the following useful formula

\be
\overline{D}^j_{nm}(x)=(-q)^{m-n}D^j_{-n,-m}(x)
\ee
\ni
which gives

\be
\int dx D^j_{n_1m_1}(x) D^{j'}_{n_2m_2}(x) = {(-q)^{m_1+n_1}\over [2j+1]}
\delta^{j,j'}\delta_{n_1,-n_2}\delta_{m_1,-m_2}
\ee
\ni
Hence, we get, instead of eq. (\ref{conjA}), the following "hermiticity"
condition

\be
\overline{A}_{\;j_1,\;j_2,\;j_3}^{m_1,m_2,m_3} = (-1)^{j_1+j_2+j_3}
(-q)^{m_1+m_2+m_3}A_{\ j_1,\ j_2,\ j_3}^{-m_1,-m_2,-m_3}
\ee
\ni

Quantum $3j$ and $6j$-symbols were investigated in ref. \cite{Kiresh},
which contains many useful formulas. The $3j$ symbol is connected to the
Clebsch-Gordan coefficient as follows

\be
\threej{j_1}{n_1}{j_2}{n_2}{j_3}{n_3}_q=
(-1)^{j_1-j_2}\frac{(-q)^{-n_3}}{\sqrt{[2j+1]}}
\langle j_1 j_2 n_1 n_2\mid j_1 j_2 j_3 -n_3 \rangle
\ee
\ni
and the eq. (\ref{int3me}) is still valid.

It is easy to see that

\be
\threej{j_1}{n_1}{j_2}{n_2}{j_3}{n_3}_q=q^{-2n_3}
\threej{j_3}{n_3}{j_1}{n_1}{j_2}{n_2}_q
\label{permsym}
\ee
\ni
And the cyclic symmetry condition in the form (\ref{cycsim}) cannot be
imposed in the quantum case.

The Racah-Wigner $6j$-symbol can be defined for example as follows

\[
\sixj{j_1}{j_2}{j_3}{j_4}{j_5}{j_6}_q=
\sum_{\{-j_i\leq m_i \leq j_i\}} (-1)^{j_4+j_5+j_6}(-q)^{m_4+m_5+m_6}
\]\[
\threej{j_5}{m_5}{j_6}{-m_6}{j_1}{m_1}_q
\threej{j_6}{m_6}{j_4}{-m_4}{j_2}{m_2}_q
\]\be
\threej{j_4}{m_4}{j_5}{-m_5}{j_3}{m_3}_q
\threej{j_1}{m_1}{j_2}{m_2}{j_3}{m_3}_q
\ee
\ni
{}From which the analog of eq. (\ref{Sintout}) immediately follows

\[
S = \frac{1}{2}\sum_{\{j_1,j_2,j_3\}}\sum_{\{-j_k \le m_k \le j_k\}}
\mid A_{\;j_1,\;j_2,\;j_3}^{m_1,m_2,m_3} \mid^2 -
\]\[
-\frac{\lambda}{4!}\sum_{\{j_1,...,j_6\}}\sum_{\{-j_k \le m_k \le j_k\}}
(-q)^{\sum_k^6 m_k}
A_{\ j_1,\ j_2,\ j_3}^{-m_1,-m_2,-m_3}
A_{\;j_3,\ j_4,\;j_5}^{m_3,-m_4,m_5} \]\be
A_{\;j_1,\ j_5,\;j_6}^{m_1,-m_5,m_6}
A_{\;j_2,\ j_6,\;j_4}^{m_2,-m_6,m_4}
(-1)^{\sum_k^6 j_k}\sixj{j_1}{j_2}{j_3}{j_4}{j_5}{j_6}_q
\ee

Now, it is strightforward to generalize eq. (\ref{Z2}) to the quantum
case. One should take care of an order of matrix elements and use the
quantum $\delta$-functions:

\be
\delta_q(x,y)=\sum_{j=0}^{M-1}[2j+1]\sum_{a,b=-j}^{j}
q^{-2b}\overline{D}^j_{ab}(x)D^j_{ab}(y)
\ee
\ni
With this definition

\be
\int dx\; f(x)\delta(x,y)=\int dx\; \delta(y,x)f(x) =f(y)
\ee

One can imagine every $\delta$-function in eq. (\ref{Z2}) as an index
loop going around a link of a triangulation. Matrices forming
the argument of the $\delta$-function can be identified with
intersections between the loop and triangles sharing the link.
In the $SU_q(2)$ case, such loops can form non-trivial knots and links
\cite{Kiresh}. If the corresponding
links\footnote{In
the knot theory sense.} are trivial, equations (\ref{2tetr}),
(\ref{3tetr}) and (\ref{pointout}) are valid in the quantum case as well
and we have the same proof of the topological invariance as for
classical groups.

A thorough investigation of the model formulated
in this section is beyond the scope of the present paper and
will be given elsewhere. A discussion on calculations of the Turaev-Viro
invariant for the lenses can be found in ref. \cite{Tata}.
In order to conclude, let us notice that this
invariant is more sensitive than the one considered in
Section 3. In principle, it can distinguish between manifolds having the
same fundamental group, which makes it, potentially, to be a powerful tool
in the theory of $3d$ manifolds.

\section{Discussion}

The models considered in this paper may be regarded as generalizing the
well-known $2d$ matrix models to the $3d$ case. They are adequate to the
problem of $3d$ euclidean quantum gravity, since they contain the sufficient
number of parameters and allow a topological expansion to be performed.

In $d$-dimensional space a metric has $d(d-1)/2$ angular degrees of
freedom which can be simulated by summing over equilateral simplicial
complexes. The other $d$ degrees of freedom are the gauge ones and one can
simply ignore them while working with a fixed topology (as in numerical
simulations in refs. \cite{AgMig,BK,4dsim}).
However, a complete theory has to
take into account both types of degrees of freedom. The aim of this
paper was to formulate such a model. Different choices of
the gauge group may be interpreted as different space structures. It
would be interesting to solve the "inverse" problem, {\em i.e.}, to
recover the geometry of a "space" (if any) corresponding to a particular gauge
group. The cyclic group $Z_n$, from this point of view, corresponds to a space
in which all lengths are quantized to be integers {\bf mod} $n$
but, instead of the triangle inequality, one has the one-dimensional
"triangle equality".
It is, in a sense, actually a model of lattice quantum gravity but
with a one-dimensional "target space".

The lattice gauge theory with a
quantum gauge group also may be of interest. It is easy to
introduce an action in it ({\em e.g.}, by eq. (\ref{heatkernel})). In this
case, the theory exists for general $q$ as well and can be
generalized to an arbitrary quantum group {\em \`{a} la}
Woronowicz. It is a theory with dynamical degrees of freedom and might
be useful in a search for new physics.

\bigskip
{\Large \bf Aknowlegements}
\medskip

The author would like to thank D.Bernard, P.Ginsparg, V.A.Kazakov,
A.Krzywicki, A.A.Migdal, M.A.Semenov-Tian-Shansky
for helpful discussions and, especially, C.Itzykson for his encouraging
interest and reading the paper
and also the Service de Physique Th\'{e}orique de Saclay for hospitality.

\pagebreak

\vspace{2cm}
\begin{center}
{\huge \bf Figures}
\end{center}

\setlength{\unitlength}{2pt}
\begin{picture}(170,100)

\thicklines
\put(10,40){\line(2,3){10}}
\put(10,40){\line(3,-1){15}}
\put(10,40){\line(1,-2){10}}
\put(20,55){\line(1,-4){5}}
\put(20,55){\line(1,-1){15}}
\put(20,20){\line(1,3){5}}
\put(20,20){\line(3,4){15}}
\put(25,35){\line(2,1){10}}
\multiput(10,40)(5,0){5}{\line(1,0){4}}

\put(40,43){\vector(1,0){10}}
\put(50,37){\vector(-1,0){10}}

\put(55,40){\line(2,3){10}}
\put(55,40){\line(3,-1){15}}
\put(55,40){\line(1,-2){10}}
\put(65,55){\line(1,-4){5}}
\put(65,55){\line(1,-1){15}}
\put(65,20){\line(1,3){5}}
\put(65,20){\line(3,4){15}}
\put(70,35){\line(2,1){10}}
\multiput(55,40)(5,0){5}{\line(1,0){4}}
\thinlines
\put(65,20){\line(0,1){35}}

\put(40,10){\makebox(10,10){\Large (a)}}

\thicklines
\put(90,35){\line(1,2){10}}
\put(90,35){\line(3,-1){15}}
\put(100,55){\line(1,-5){5}}
\put(100,55){\line(3,-4){15}}
\put(105,30){\line(2,1){10}}
\multiput(90,35)(5,0){5}{\line(1,0){4}}

\put(120,43){\vector(1,0){10}}
\put(130,37){\vector(-1,0){10}}

\put(140,35){\line(1,2){10}}
\put(140,35){\line(3,-1){15}}
\put(150,55){\line(1,-5){5}}
\put(150,55){\line(3,-4){15}}
\put(155,30){\line(2,1){10}}
\multiput(140,35)(5,0){5}{\line(1,0){4}}
\thinlines
\put(150,40){\line(0,1){15}}
\put(150,40){\line(3,-1){15}}
\put(150,40){\line(1,-2){5}}
\put(150,40){\line(-2,-1){10}}

\put(120,10){\makebox(10,10){\Large (b)}}

\end{picture}

\begin{description}
   \item[Fig. 1] ---
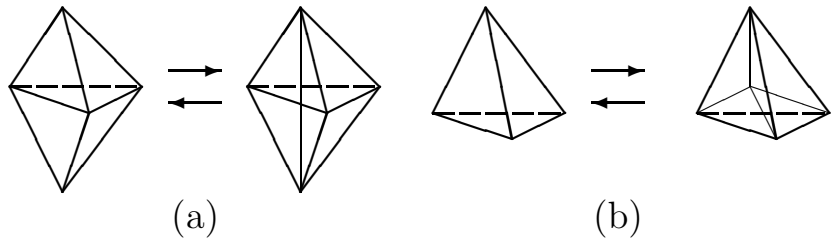
(a) The triangle-link exchange: the common triangle of two tetrahedra on
the left is removed and three new triangles sharing the new link appear
on the right.
(b) The subdivision: 4 new tetrahedra fill an old one.
\end{description}

\begin{picture}(170,100)
\thicklines

\put(30,30){\line(0,1){20}}
\put(30,30){\line(2,-1){10}}
\put(30,30){\line(1,-2){5}}
\put(30,30){\line(-2,-1){10}}
\put(30,30){\circle*{2}}
\put(30,50){\line(2,1){10}}
\put(30,50){\line(1,2){5}}
\put(30,50){\line(-2,1){10}}
\put(30,50){\circle*{2}}

\put(40,43){\vector(1,0){10}}
\put(50,37){\vector(-1,0){10}}

\put(60,40){\circle*{2}}
\put(60,40){\line(-1,3){5}}
\put(60,40){\line(-1,-3){5}}
\put(60,40){\line(1,0){10}}
\put(60,40){\line(2,-1){10}}
\put(70,35){\circle*{2}}
\put(70,35){\line(1,5){5}}
\put(70,35){\line(1,-3){5}}
\put(80,40){\circle*{2}}
\put(80,40){\line(1,3){5}}
\put(80,40){\line(1,-3){5}}
\put(80,40){\line(-2,-1){10}}
\put(80,40){\line(-1,0){8}}

\put(40,5){\makebox(10,10){\Large (a)}}

\put(110,25){\line(0,1){30}}

\put(120,43){\vector(1,0){10}}
\put(130,37){\vector(-1,0){10}}

\put(145,25){\line(0,1){30}}
\put(145,40){\oval(12,16)}
\put(145,32){\circle*{2}}
\put(145,48){\circle*{2}}

\put(120,5){\makebox(10,10){\Large (b)}}

\end{picture}

\begin{description}
   \item[Fig. 2] ---
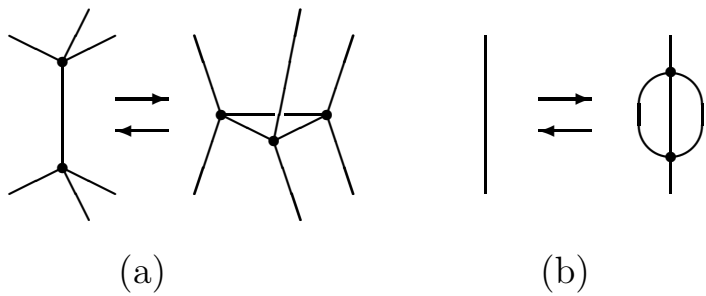
Dual graphs: (a) the triangle-link exchange; (b) two tetrahedra glued
along three common faces (a self-energy insertion) are equivalent to a
triangle.
\end{description}


\begin{picture}(170,100)
\thicklines

\put(5,40){\circle*{2}}
\put(5,40){\line(1,0){15}}
\put(20,40){\circle*{2}}

\put(25,40){\vector(1,0){10}}

\put(40,40){\circle*{2}}

\put(0,40){\makebox(10,10){$\sigma^0_1$}}
\put(15,40){\makebox(10,10){$\sigma^0_2$}}
\put(35,40){\makebox(10,10){$\sigma^0$}}

\put(15,5){\makebox(10,10){\Large (a)}}

\put(70,40){\circle{25}}
\put(70,40){\oval(20,8)}
\multiput(62,37)(2,0){6}{\line(1,1){6}}
\put(74,37){\line(1,1){5}}
\put(76,37){\line(1,1){3}}
\put(60,38){\line(1,1){5}}
\put(58,39){\line(1,1){3}}
\put(65,32){\makebox(10,10)[b]{$\sigma^{\scs{2}}_{\scs{1}}$}}

\put(65,46){\makebox(10,10)[b]{$\sigma^{\scs{2}}_{\scs{2}}$}}

\put(85,40){\vector(1,0){10}}
\put(110,40){\circle{25}}
\put(105,35){\makebox(10,10){$\sigma^2$}}

\put(85,5){\makebox(10,10){\Large (b)}}

\put(135,40){\circle*{15}}
\put(145,40){\vector(1,0){10}}
\put(160,40){\circle*{2}}
\put(155,40){\makebox(10,10){$\sigma^0$}}
\put(125,45){\makebox(20,10){$S^3\backslash(\bullet)$}}

\put(145,5){\makebox(10,10){\Large (c)}}

\end{picture}

\begin{description}
   \item[Fig. 3] ---
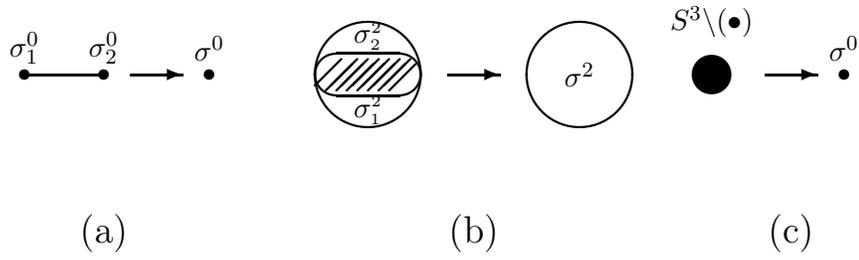
A subcomplex can be substituted by a homotopic one: (a)
$\sigma^0_1=\sigma^0_2=\sigma^0$; (b) $\sigma^2_1\cup\sigma^2_2=\sigma^2$;
(c) a $3d$ ball is homotopic to a point.
\end{description}

\pagebreak


\begin{thebibliography}{99}

\bibitem{MatMod} V.A.Kazakov, Phys.Lett {\bf B150} (1985) 282;
F. David, Nucl. Phys. {\bf B257} (1985) 45,543;
J. Ambj\o rn, B.Durhuus and J. Fr\"{o}hlich
Nucl. Phys. B {\bf 257} (1985) 433;
 D.V.Boulatov, V.A. Kazakov, I.K. Kostov and A.A. Migdal, Nucl.Phys.
 {\bf B275} (1986) 641.

\bibitem{Regge} T.Regge, Nouvo Cimento {\bf 19} (1961) 558.

\bibitem{AgMig} M.E. Agishtein and A.A. Migdal, Mod. Phys. Lett.
{\bf A6} (1991) 1863;
J.Ambj\o rn and S. Varsted,Phys.Lett. {\bf B226} (1991) 258.

\bibitem{BK} D.V. Boulatov and A. Krzywicki, Mod. Phys. Lett. {\bf A6}
(1991) 3005;
J.Ambj\o rn, D.Boulatov, A.Krzywicki and S.Varsted, {\it The vacuum in
three-dimensional  simplicial quantum gravity"}, preprint NBI-HE-91-46,
LPTHE Orsay 91/57 (October 1991).

\bibitem{4dsim} M.E. Agishtein and A.A. Migdal, {\it Simulations of
four-dimensional simplicial quantum gravity"}, preprint PUPT-1287
(October 1991);
J.Ambj\o rn and J.Jurkiewicz, preprint NBI-HE-91-47 (November 1991).

\bibitem{TensorMod} D.V.Boulatov and V.A.Kazakov, unpublished;
 J. Ambj\o rn, B. Durhuus and T. Jonsson,
Mod. Phys. Lett. {\bf A6} (1991) 1133;
N.Sasakura, Mod. Phys. Lett. {\bf A6} (1991) 2613;
N.Godfrey and M.Gross, Phys. Rev. {\bf D43} (1991) R1749.

\bibitem{Matmod2} E.Brezin and V.A.Kazakov, Phys.Lett. {\bf B236} (1990)
144;
M.Douglas and S.Shenker, Nucl. Phys. {\bf B335} (1990) 635;
D.Gross and A.A.Migdal, Phys. Rev. Lett. {\bf 64} (1990) 127.

\bibitem{Witten} E.Witten, Commun.Math.Phys. {\bf 121} (1989) 351;
Nucl. Phys. {\bf B311} (1988/89) 46.

\bibitem{TV} V.G.Turaev and O.Y.Viro, {\it State sum invariants of 3-manifolds
and quantum $6j$-symbols"}, LOMI preprint (1990).


\bibitem{Ooguri} H.Ooguri and N.Sasakura, Mod. Phys. Lett. {\bf A6}
(1991) 3591;
F.Archer and R.M.Williams, Phys.Lett. {\bf B273} (1991) 438.

\bibitem{PR} G.Ponzano and T. Regge, {\it in Spectroscopic and group
theoretical methods in physics}, ed. F.Bloch (North-Holland, Amsterdam,
1968).

\bibitem{DijkW} R.Dijkgraaf and E.Witten, Commun. Math. Phys. {\bf 129}
(1990) 393.

\bibitem{Woron} S.L.Woronowicz, Commun. Math. Phys. {\bf 111}
(1987) 613;
Ya.Soibelman, Algebra i analiz {\bf 2} (1990) 190.

\bibitem{Majid} S.Majid, Int. J. Mod. Phys. {\bf A 5} (1990) 1.

\bibitem{repsu2} E.K.Sklyanin, Funct. Anal. Appl. {\bf 17} (1983) 273;
P.Roche and D.Arnaudon, Lett. Math. Phys. {\bf 17} (1989) 295.

\bibitem{Keller} G.Keller, Lett. Math. Phys. {\bf 21} (1991) 273.

\bibitem{Kiresh} A.N.Kirillov and N.Yu.Reshetikhin, {\em Representations
of the algebra ${\cal U}_q(SL_2)$, q-orthogonal polynomials and
invariants of links}, LOMI preprint E-9-88, Leningrad 1988.

\bibitem{Tata}
S.Rama and S.Sen, {\em "Three manifolds and graph invariants"},
Tata Institute preprint TIFR/TH/91-59 (December
1991).

\end{thebibliography}
\end{document}